\documentclass[usenatbib,a4paper,times,fleqn]{mnras}
\usepackage{graphicx,natbib,times,amssymb,amsmath,pstricks}
\usepackage{aas_macros}
\usepackage{bm,url}
\usepackage{textcomp}
\usepackage{breakurl}

\title[Dust and gas in HL Tau]{On planet formation in HL Tau}

\author[Dipierro et al.]{\parbox{\textwidth}{Giovanni Dipierro$^{1,2}$\thanks{giovanni.dipierro@unimi.it}, Daniel Price$^{2}$\thanks{daniel.price@monash.edu}, Guillaume Laibe$^{3}$, Kieran Hirsh$^{2}$, Alice Cerioli$^{1}$ and Giuseppe Lodato$^{1}$} \vspace{0.1cm} \\
$^{1}$Dipartimento di Fisica, Universit\`a Degli Studi di Milano, Via Celoria, 16, Milano, I-20133, Italy \\
$^{2}$Monash Centre for Astrophysics (MoCA) and School of Physics and Astronomy, Monash University, Clayton Vic 3800, Australia\\
$^{3}$School of Physics and Astronomy, University of St. Andrews, North Haugh, St. Andrews, Fife KY16 9SS, UK
}
\pagerange{\pageref{firstpage}--\pageref{lastpage}} \pubyear{2015}

\date{}
\begin{document}
\label{firstpage}
\bibliographystyle{mnras}
\maketitle

\begin{abstract}
 We explain the axisymmetric gaps seen in recent long-baseline observations of the HL Tau protoplanetary disc with the Atacama Large Millimetre/Submillimetre Array (ALMA) as being due to the different response of gas and dust to embedded planets in protoplanetary discs. We perform global, three dimensional dusty smoothed particle hydrodynamics calculations of multiple planets embedded in dust/gas discs which successfully reproduce most of the structures seen in the ALMA image. We find a best match to the observations using three embedded planets with masses of 0.2, 0.27 and 0.55 $M_{\rm J}$ in the three main gaps observed by ALMA, though there remain uncertainties in the exact planet masses from the disc model.
\end{abstract}


\section{Introduction}
 Recent long-baseline observations of the HL Tau protoplanetary disc at mm to cm wavelengths using the Atacama Large Millimetre/Submillimetre Array (ALMA) revealed a striking series of concentric, axisymmetric gaps, thought to indicate the presence of massive protoplanetary bodies caught in the act  of forming planets \citep{almaetal15}. 
 
While gap carving by planets embedded in discs has been predicted theoretically since at least \citet{linpapaloizou86}, the axisymmetry of the rings in HL Tau is puzzling. First, prominent spiral waves induced by the planet are seen in all theoretical simulations of planets carving gaps in gas discs whenever the planet is massive enough to open a gap \citep[e.g.][]{de-val-borroetal06}. Second, prior observational estimates of the relative disc mass in HL Tau suggested a disc of 0.03--0.14 $M_{\odot}$ around a 0.5-1.3 $M_{\odot}$ star (e.g. \citealt{robitailleetal07,klm11,kwonetal15}; the latter authors estimated $\approx 0.1$ M$_{\odot}$ based on fitting disc models to the previous best observations at mm wavelengths). Massive discs should almost certainly show gravitationally instability and hence prominent and large scale spiral arms \citep[e.g.][]{lodatorice04}, as seen in the theoretical predictions for HL Tau made by \citet{greavesetal08}. Third, HL Tau is in an early phase of star formation (Class I evolving to Class II), emitting a 100-200 km/s highly collimated jet \citep{mundtetal90}. The ongoing accretion from the envelope means that the disc is likely relatively thick, massive and hot, yet the clarity of the gaps are more reminiscent of those found in simulations of thin, cold discs (e.g. our most successful attempt to reproduce the gaps in gas-only discs required $H/R \lesssim 0.02$).

This begs the question of whether the gaps seen in HL Tau are genuine signatures of planet formation, or due to some other phenomena such as clumping \citep{lyrakuchner13}, magnetic or Rossby-wave instabilities \citep{pinillaetal12} or condensation fronts \citep{zbb15}. Then again, the observation that the first four dark rings are located close to what would be resonances between planets orbiting in those locations, together with slight eccentricity observed in the rings are evidence in favour of carving by planets \citep{wolfetal02,almaetal15}.

  Key is that the emission seen at the wavelengths observed by ALMA is from dust rather than gas. Dust grains settle and migrate in protoplanetary discs at a rate determined by the Stokes number, $S_{\rm t}$, the ratio of the stopping time to the orbital timescale \citep{weidenschilling77,takeuchilin02} with the stopping time depending on the grain size and gas density (Eq.~\ref{eq:ts}). If mm-grains have settled to the mid-plane in HL Tau then the gap-carving seen by ALMA could be that induced in the thin dust disc rather than the relatively thicker gas disc, since it is easier to carve gaps in the dust \citep{paardekoopermellema04}. The absence of spiral structure may also be explained by the dust. Simulations of dust grains near gaps carved by planets \citep{paardekoopermellema04,paardekoopermellema06,mfg07,fouchetetal07,fgm10,ayliffeetal12} have demonstrated that dust grains close to $S_{\rm t}=1$ form axisymmetric rings, with smaller dust grains more closely following the gas and larger grains trapped in resonances \citep{ayliffeetal12}.
 
 Whether embedded planets in dusty gas discs can explain the axisymmetric gaps in HL Tau is the focus of this Letter.
 

\section{Methods}
\label{sec:methods}

\subsection{Dust/gas simulations}
We perform 3D global simulations of dust/gas discs with embedded protoplanets using the \textsc{phantom} smoothed particle hydrodynamics (SPH) code written by DP \citep*{pricefederrath10,lodatoprice10,price12,nkp13}.

 We assume a single grain size per calculation. For grains of mm-size and larger we employ the two fluid algorithm described in \citet{laibeprice12,laibeprice12a}. For calculations with smaller grains we employ our new single fluid algorithm based on the terminal velocity approximation \citep{laibeprice14,pricelaibe15}. This algorithm has been extensively benchmarked on simple test problems including waves and shocks in dust/gas mixtures \citep{laibeprice11,laibeprice12,pricelaibe15}, but this is our first application to planet formation, allowing us to evolve both small and large grains efficiently.

\subsubsection{Drag prescription}
We assume a physical drag prescription where the drag regime is chosen automatically depending on the ratio of the grain size to the gas mean free path \citep{kwok75,paardekoopermellema06,laibeprice12a}. All of the grain sizes used in this paper fall into the Epstein regime, meaning that the stopping time is given by
\begin{equation}
t_{\rm s} = \frac{ \rho_{\rm grain} s_{\rm grain}}{\rho c_{\rm s}f}  \sqrt{\frac{\pi\gamma}{8}},
\label{eq:ts}
\end{equation}
where $\rho_{\rm grain}$ is the intrinsic grain density (we assume 1 g/cm$^{3}$), $s_{\rm grain}$ is the grain size, $\rho = \rho_{\rm g} + \rho_{\rm d}$ is the total density of the mixture, $c_{\rm s}$ is the sound speed, $\gamma = 1$ is the adiabatic index and $f$ is a correction factor for supersonic drag given by \citep{kwok75}
\begin{equation}
f \equiv \sqrt{1 + \frac{9\pi}{128} \frac{\Delta v^{2}}{c_{\rm s}^{2}}},
\end{equation}
where $\Delta v \equiv \bm{v}_{\rm d} - \bm{v}_{\rm g}$ is the drift velocity between the dust and gas. Hence, the Stokes number $S_{\rm t} \equiv t_{\rm s}\Omega$ in the mid-plane is
\begin{equation}
S_{\rm t} = 1 \bigg( \frac{\Sigma}{0.2 {\rm g}/{\rm cm}^{2}} \bigg)^{-1} \bigg( \frac{\rho_{\rm grain}}{3 {\rm g}/{\rm cm}^{3}} \bigg) \bigg( \frac{s_{\rm grain}}{1 {\rm mm}} \bigg) \bigg(\frac{f}{1} \bigg)^{-1}.
\label{eq:stokes}
\end{equation}
A caveat to the above drag prescription is that it assumes spherical grains. Fractal grain shapes or similar would change the effective Stokes number associated with a particular grain size.

\subsubsection{Sink particles}
 We place three planets, initially located at 13.2, 32.3 and 68.8 au, in the three main gaps observed in HL Tau \citep{almaetal15}. We represent the protoplanetary bodies and the central star using sink particles \citep*{bbp95} with accretion radii of 0.25, 0.25 and 0.75 au, respectively. Dust and gas can be accreted onto the sinks provided that the material is bound to the sink and the velocity divergence is negative. The planets are free to migrate due to planet-disc interactions and the viscous disc evolution. We found migration to be negligible on the timescales simulated.
 

\subsection{Initial conditions}

\subsubsection{Gas}
We set up the discs in \textsc{phantom} using the standard procedure outlined in \citet{lodatoprice10}. 
The system comprises a central star with mass $1.3 M_{\odot}$ surrounded by a gaseous disc of $10^{6}$ SPH particles which extends from $R_{\rm{in}}$ = 1 to $R_{\rm{out}}$ = 120 au.  We assume a power-law surface density profile given by 
$\Sigma(R) = \Sigma_{\rm in} (R/R_{\rm in})^{-p}$,
where $\Sigma_{\rm in}$ is the surface density at the inner edge. We assume a radially isothermal equation of state $P=c_{\rm s}^{2}(R)  \rho$, where 
$c_{\rm s} = c_{{\rm s,in}} (R/R_{\rm in})^{-q}$, such that the gas temperature power law is $T(R) \propto (R/R_{\rm in})^{-2q}$ and where $c_{{\rm s},{\rm in}}$ is the sound speed at $R_{\rm in}$. The aspect ratio in the gas is thus prescribed by the $p$ and $q$ indices according to
$H/R = (H/R)_{\rm in} \,(R/R_{\rm in})^{\frac12 - q}$.
 We set the SPH $\alpha_{\rm AV}$ viscosity parameter to 0.1 giving an effective \citet{shakurasunyaev73} viscosity $\alpha_{SS} \approx 0.005$.

\begin{figure}
\begin{center}
\vspace{-0.3cm}
\includegraphics[width=\columnwidth]{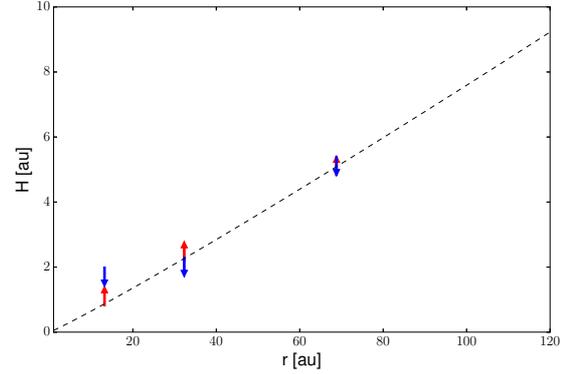}
\caption{Constraint on the disc scale height profile obtained from the gap widths observed in the ALMA image (upper limits; blue points) combined with the lower limit obtained from the dust temperature as a function of radius (red points) along with our best fit $T \propto R^{-0.7}$ (dashed line).} 
\label{fig:HonR}
\end{center}
\end{figure}

\subsubsection{Constraining the disc model from HL Tau}
While \citet{klm11} fit various disc profiles to their previous low-resolution observations of HL Tau to constrain the surface density and temperature profiles, we use the gaps observed in the disc \citep{almaetal15} to directly constrain the aspect ratio.

 In particular, while the gap is deeper in the dust \citep{fouchetetal07}, the \emph{width} is the same. Since a gap cannot be opened with a width less than the disc scale height the three prominent gaps observed in HL Tau provide an upper limit on $H$ as a function of radius. Combining this with the dust temperature profile observed by ALMA, which, assuming that the dust and gas are thermally coupled, provides a lower limit on the gas temperature, we can directly constrain the sound speed profile. Figure~\ref{fig:HonR} shows the constraint provided by the gap widths together with the lower limit and our best fitting power law $q=0.35$ ($T \propto R^{-0.7}$) with $(H/R)_{\rm in} = 0.04$. This gives $(H/R)_{\rm out} \approx 0.08$.
 
 We choose a shallow surface density profile $p= 0.1$ for computational efficiency since it avoids expensive computation of many particles on the shortest orbits. This reproduces the inner disc in our simulations, but gives an outer disc that is over-luminous compared to the inner disc in the simulated observations (Fig.~\ref{fig:cfr}). While more work is needed on the disc model, our main aim in the present paper is to reproduce the axisymmetry and size of the gaps.
 
\subsubsection{Gas disc mass and dust-to-gas ratio}
 \citet{hom93} found a H$_{2}$ gas mass of 0.03$M_{\odot}$ contained within 1400 au in HL Tau based on the observed $^{13}$CO emission. We assume this as our starting point, noting the the mass contained with a particular radius depends on the density profile according to
\begin{equation}
M(<R) = \frac{2\pi}{2- p} \Sigma_{\rm in} R_{\rm in}^{2} \left[\left(\frac{R}{R_{\rm in}}\right)^{2-p} - 1 \right].
\end{equation}
 We assume the dust was initially co-located with the gas but has migrated to a radius $\approx$ 10 times smaller than the initial radius to give the 120 au dust disc visible with ALMA. Assuming an initial dust-to-gas ratio of 0.01 in the undifferentiated material at 1400 au, this gives $\Sigma_{\rm gas} \approx \Sigma_{\rm dust}$ within the smaller radius given our assumed $p$ index of 0.1. Hence we assume a dust-to-gas ratio of unity within 120 au and set $\Sigma_{\rm in}$ such that $M_{\rm disc} = 0.03 M_{\odot}$ at 1400 au. The total gas mass within 120 au is thus 0.0002 M$_{\odot}$, but this is low mainly due to our assumed shallow surface density profile. 
 
  A steeper surface density profile, e.g. $p=1$ would give a more realistic gas mass of $0.003 M_{\odot}$. While even this may seem in conflict with earlier fits to much higher disc masses \citep[e.g.][]{robitailleetal07,klm11,kwonetal15}, these assume that the dust is co-located with the gas, whereas we expect settling and radial inward migration to produce a more compact dust disc, consistent with observations in other systems \citep{andrewsetal12,de-gregorio-monsalvoetal13}. Moreover, \citet{tamayoetal15} found that such a massive gas disc within 120 au would smear out the eccentric features observed in HL Tau since the precession periods of pericentres would be faster than the timescale on which planets carve gaps. Our lower disc mass also ensures that the mm grains are close to $S_{\rm t} = 1$ in order to guarantee an efficient migration towards the pressure maxima in the disc induced by the embedded protoplanets. Again, a steeper surface density profile would produce less coupling of the mm grains in the outer disc, giving dynamics similar to those we find with cm grains. Indeed, as we shall see, the spiral waves induced in larger grains by the outer planet do appear to be visible at mm wavelengths in HL Tau.
  


\begin{figure}
\begin{center}
\includegraphics[width=0.4\textwidth]{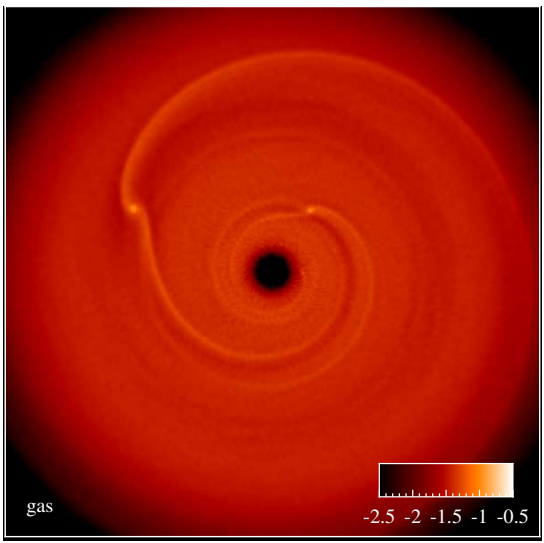}
\caption{Rendered image of gas surface density for a disc containing three embedded protoplanets of mass 0.2, 0.27 and 0.55 $M_{\rm J}$ initially located at the same distance as the gaps detected in HLTau, at 13.2, 32.3 and 68.8 au. } 
\label{fig:dumpgas}
\end{center}
\end{figure}

\begin{figure*}
\begin{center}
\includegraphics[width=0.28\textwidth]{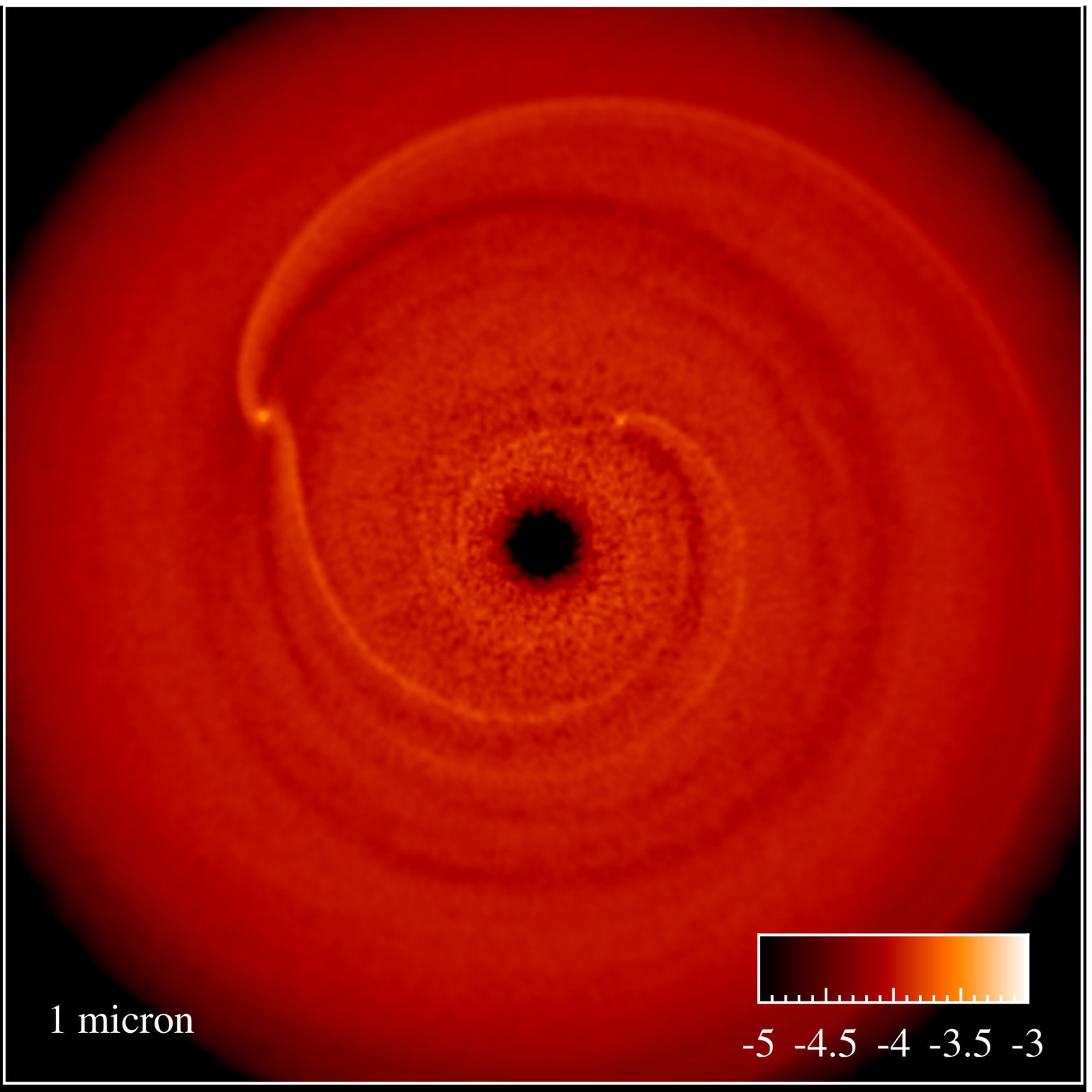}
\includegraphics[width=0.28\textwidth]{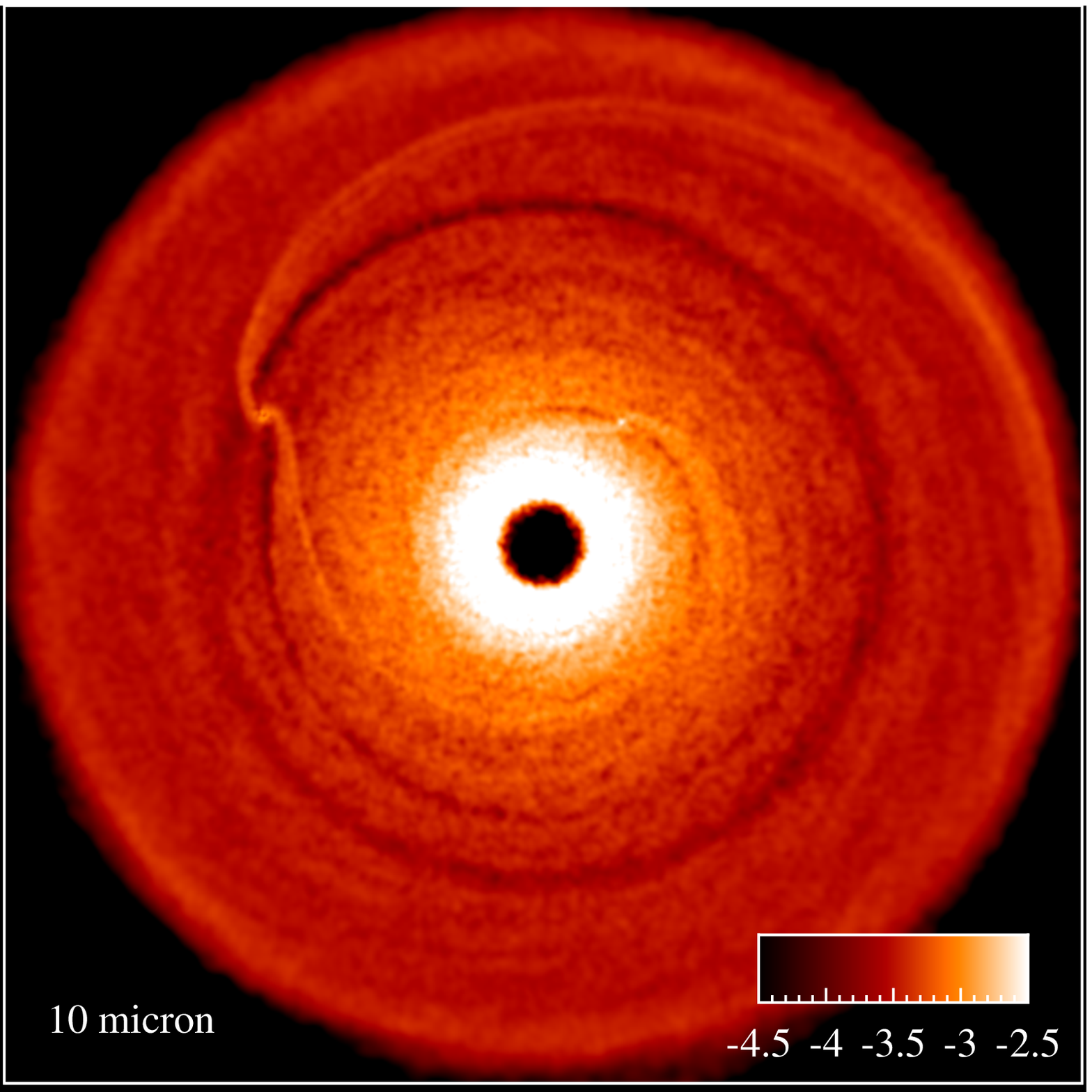}
\includegraphics[width=0.28\textwidth]{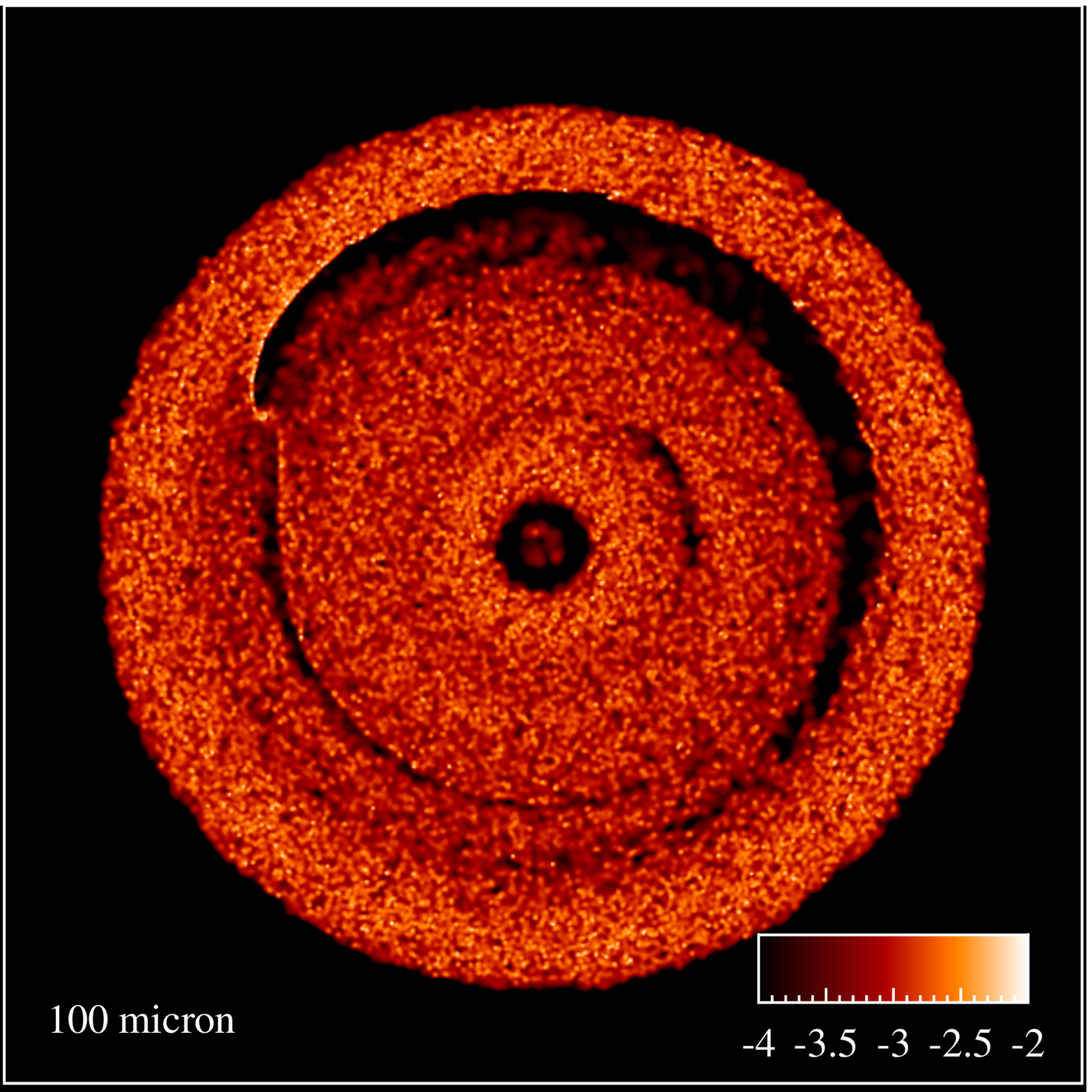}
\includegraphics[width=0.28\textwidth]{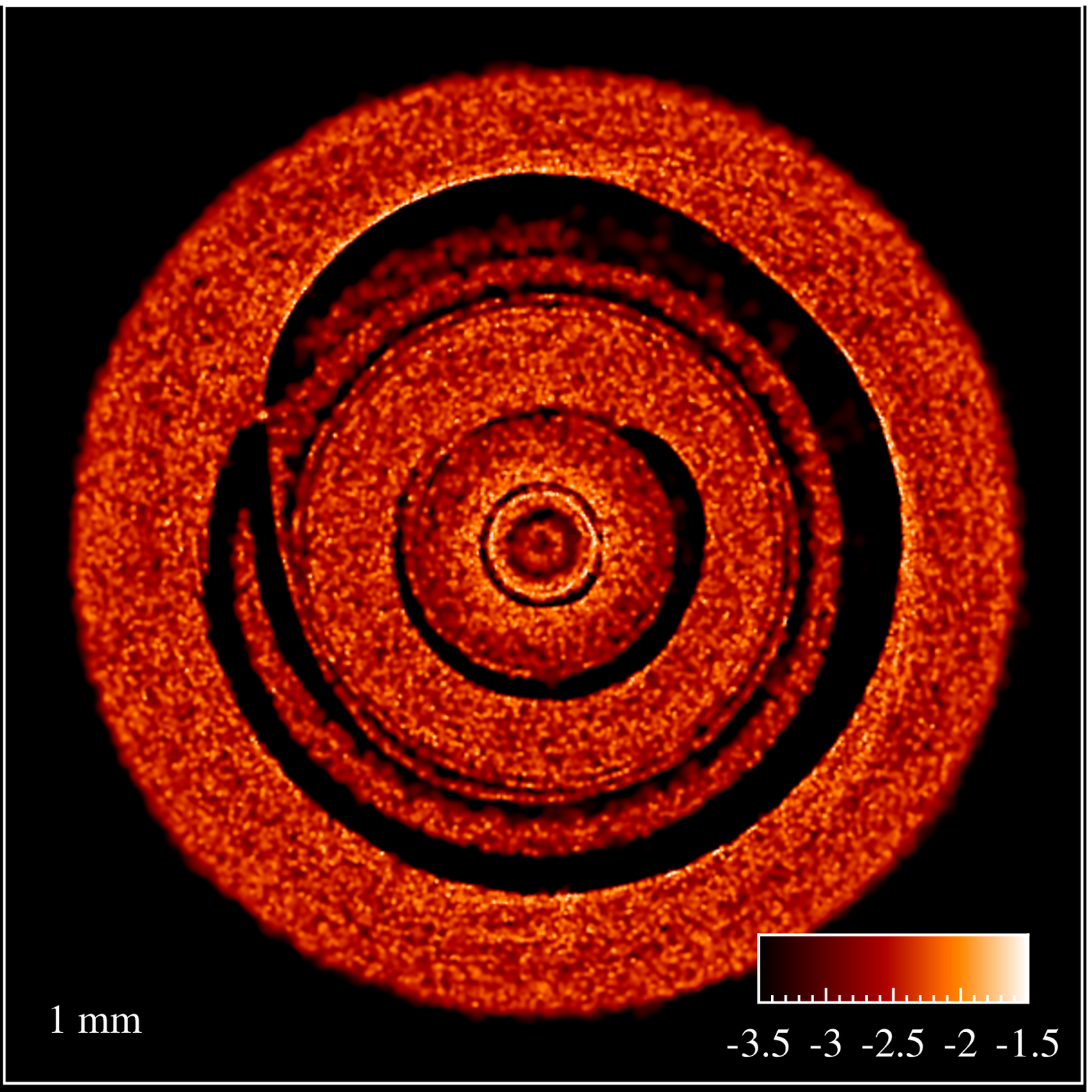}
\includegraphics[width=0.28\textwidth]{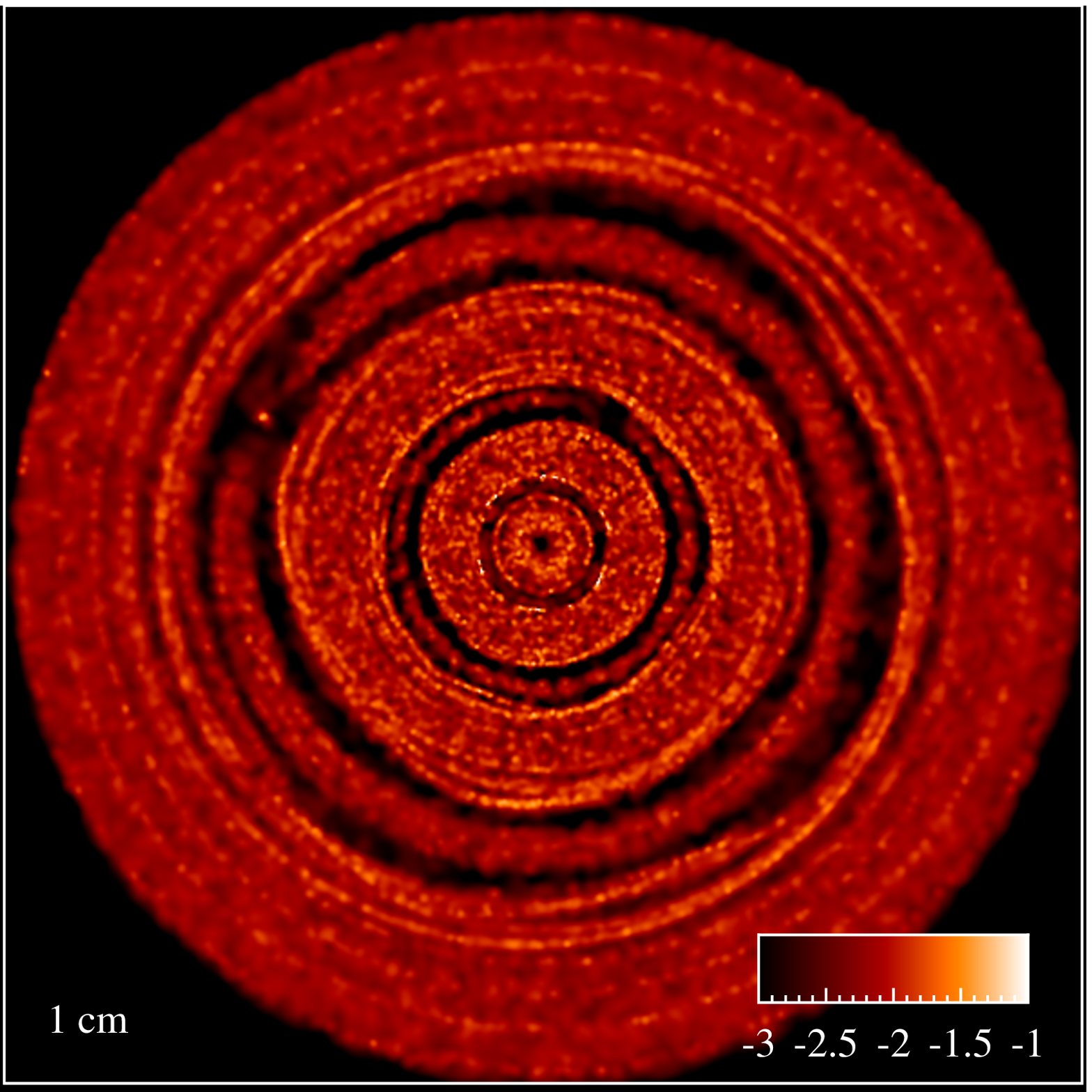}
\includegraphics[width=0.28\textwidth]{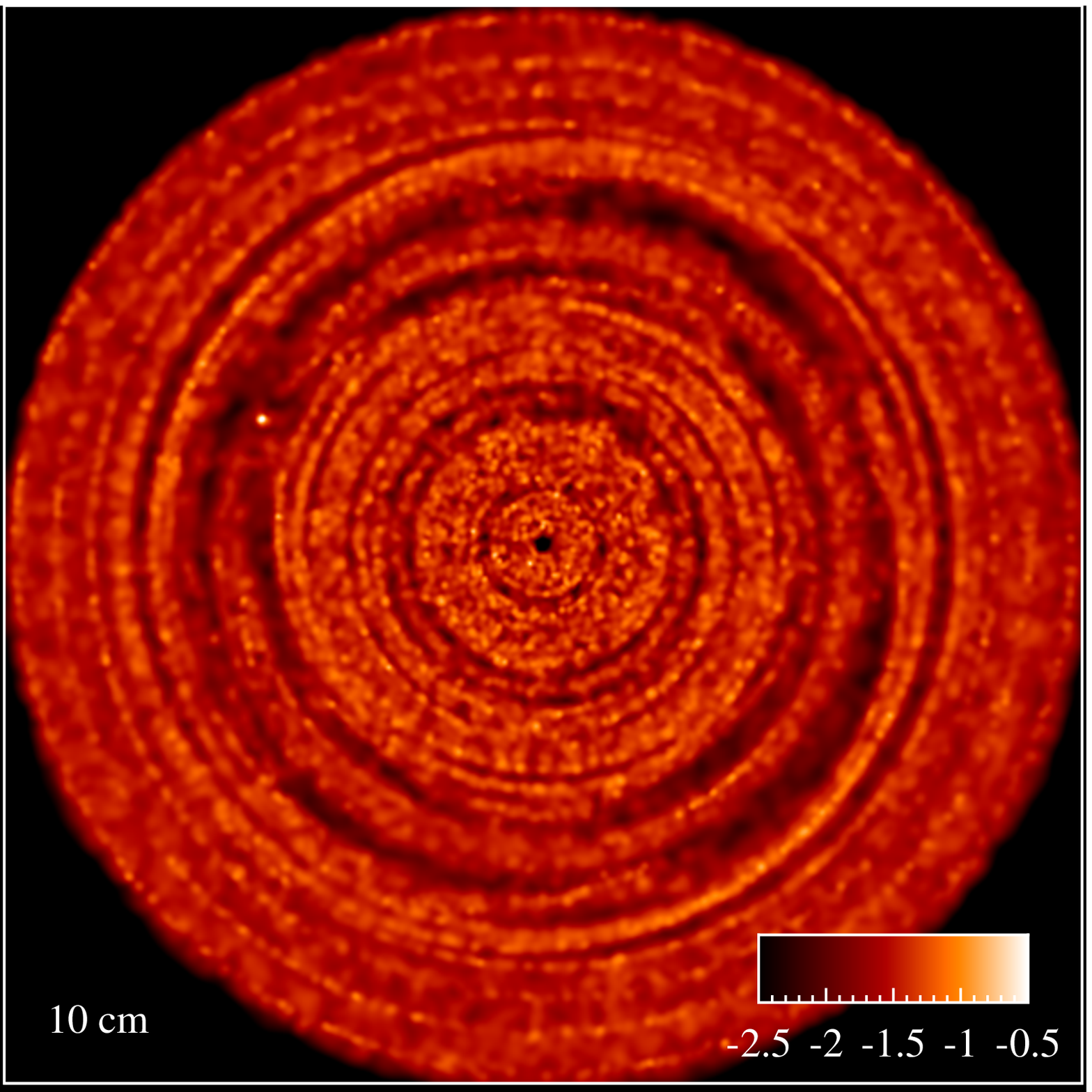}
\caption{Rendered images of dust surface density for a disc containing three embedded protoplanets of mass 0.2, 0.27 and 0.55 $M_{\rm J}$ initially located at the same distance as the gaps detected in HL Tau. Each panel shows the simulation with gas plus grains of a particular size (as indicated).}
\label{fig:dust}
\end{center}
\end{figure*}

\subsubsection{Dust}
The dust disc is setup assuming an initially constant dust-to-gas ratio (equal to unity, see above). Since we simulate one grain size at a time we assume a power-law grain size distribution given by
\begin{equation}
n(s) \propto s^{-m} \hspace{0.5cm} \textrm{for} \hspace{0.5cm} s_{\min} < s < s_{\max},
\end{equation}
where we assume $s_{\min} = 1\micron$, $s_{\max} = 10$cm and $m = 3.5$ and perform six different simulations using $s = 1\micron$, $10\micron$, $100\micron$, $1\rm{mm}$, $1\rm{cm}$ and $10\rm{cm}$, respectively (Fig.~\ref{fig:dust}). The grain size distribution is appropriate to interstellar dust and particles growing by agglomeration in protoplanetary discs \citep{draine06}. While approximate, this is more realistic than assuming a constant dust-to-gas ratio for each grain size, capturing the main idea that small grains are more abundant than large grains.


 In our two fluid simulations we set up the dust disc as a separate set of dust particles. We use one quarter of the number of dust particles compared to gas to prevent dust from becoming artificially trapped below the gas resolution \citep{laibeprice12a,ayliffeetal12}. For the calculations employing the single fluid algorithm \citep{pricelaibe15} there is only one set of particles describing the mixture, set up as described above for the gas but with a dust fraction $\epsilon$ set for each particle to provide an initially constant dust-to-gas ratio (i.e. $\epsilon = \rho_{\rm d}/\rho_{\rm g} / [1 + \rho_{\rm d}/\rho_{\rm g}]$). With both methods the dust is then free to settle and migrate in the disc, and feels gravity from both the central star and the embedded protoplanets.

\subsection{Simulated ALMA observations}
 We perform simulated observations of our model using the RADMC-3D Monte Carlo radiative transfer code \citep{dullemond12} together with the Common Astronomy Software Application (CASA) ALMA simulator (version 4.1), focussing on ALMA band 6 (continuum emission at 233 GHz). The source is located in the position of HL Tau, adopting the disc inclination and position angle given by \citet{almaetal15}.  
 The procedure is as in \citet{dipierroetal15} except that the dust distribution is used directly from our simulation data rather than being prescribed. The dust model consists of spherical silicate grains with optical constant for magnesium-iron grains taken from the Jena database\footnote{\burl{http://www.astro.uni-jena.de/Laboratory/Database/databases.html}}. 
 Full-resolution images produced by RADMC-3D simulations are used as input sky models to simulate realistic ALMA observations accounting for thermal noise from the receivers and the atmosphere and assuming a perfect calibration of the visibility measurements. 
 A transit duration of 8 hours is used to reach an optimal signal-to-noise ratio. The observation parameters are chosen to match the spatial resolution of the observations (see caption of Fig. \ref{fig:cfr}).

\section{Results}
\label{sec:results}
Fig. \ref{fig:dumpgas} shows the gas surface density, while Fig. \ref{fig:dust} shows rendered images of the dust surface density from the disc model with three embedded protoplanets of masses of 0.2, 0.27 and 0.55 $M_{\rm J}$, respectively, shown after $\sim$10 orbits of the outer planet. None of the planets produce clean gaps in the gas disc (Fig.~\ref{fig:dumpgas}), but as expected, the dust density perturbations induced by protoplanets depend strongly on the grain size \citep[e.g.][]{fouchetetal07,ayliffeetal12}, or more specifically on the Stokes number. 
The density distribution of $\micron$-sized particles (top left and centre) are similar to the gas distribution due to the stronger coupling. The micrometer sized grains (top left) capture the spiral density wave launched by the protoplanets. 
Millimetre-sized particles (bottom left) are most affected by the density waves induced by the protoplanets, exhibiting the largest migration towards the gap edges. Interestingly, the dust density distributions of 10~mm and cm-sized particles (bottom centre and right) show axisymmetric waves launched by the planets propagating across the whole disc. The gaps carved by protoplanets in cm grains ($S_{\rm t} \approx 10$) show the formation of horseshoe regions. As expected, the formation of the horseshoe region in planetary gaps depends on the dust-gas coupling, since poorly coupled large particles tend to have frequent close encounters with the planet. We stress that the important parameter is the Stokes number rather than the actual grain size, so equivalent dynamics can be produced in different grains by changing the gas density or with different assumptions about the grain composition (see Eq.~\ref{eq:stokes}).

An intriguing possibility is to infer the disc viscosity from the morphology of the gaps. Since our results show that the planets are not massive enough to open gaps in the gas or, equivalently, the viscosity is high enough to effectively smooth the density profile, we can only evaluate a lower limit for $\alpha_{SS}$ below which planets are able to carve gaps in the disc model adopted here. Using the criterion in \citet{cmm06} we infer $\alpha_{SS} \gtrsim 0.0002$.

\begin{figure*}
\begin{center}
\includegraphics[width=0.455\textwidth]{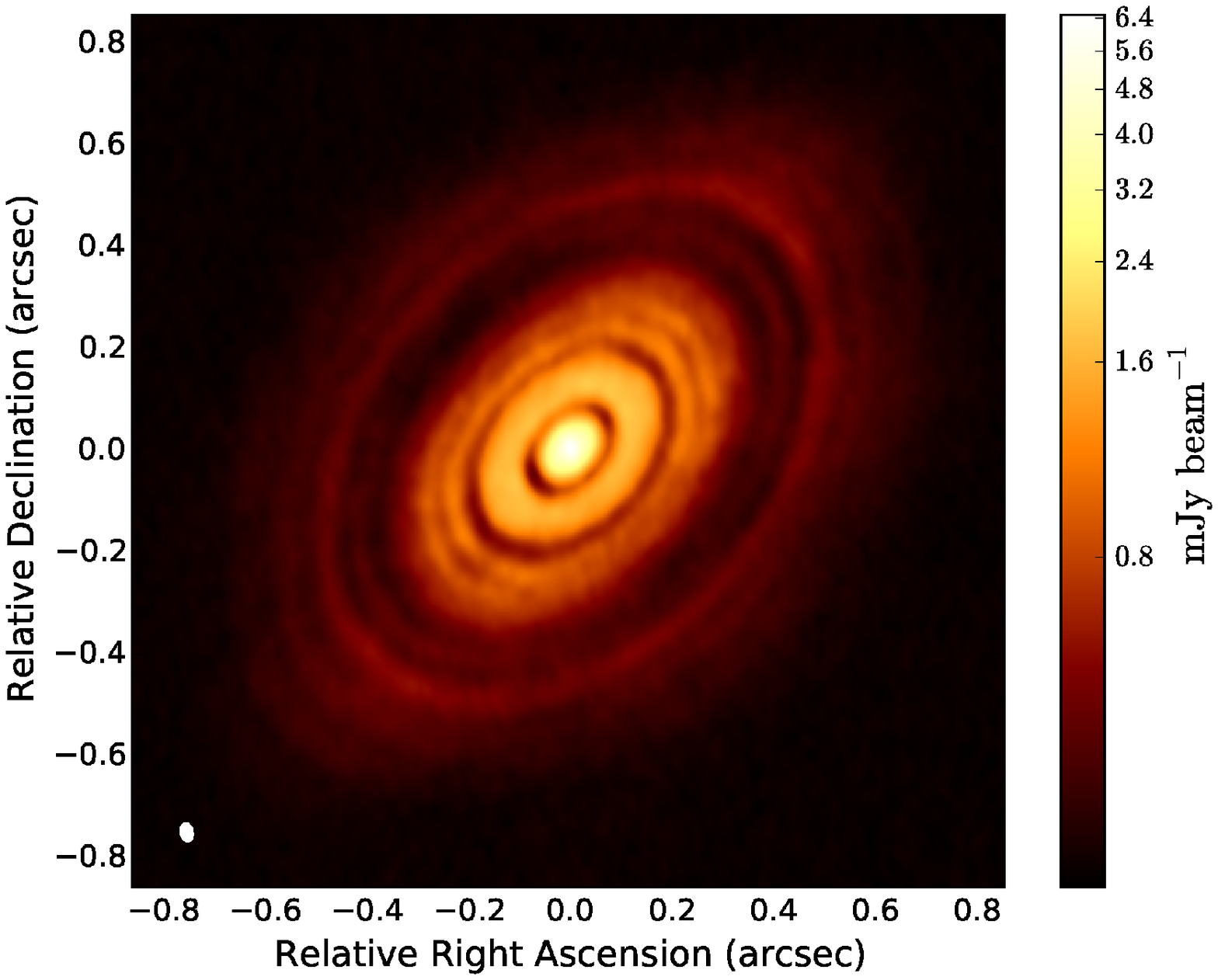}
\includegraphics[width=0.455\textwidth]{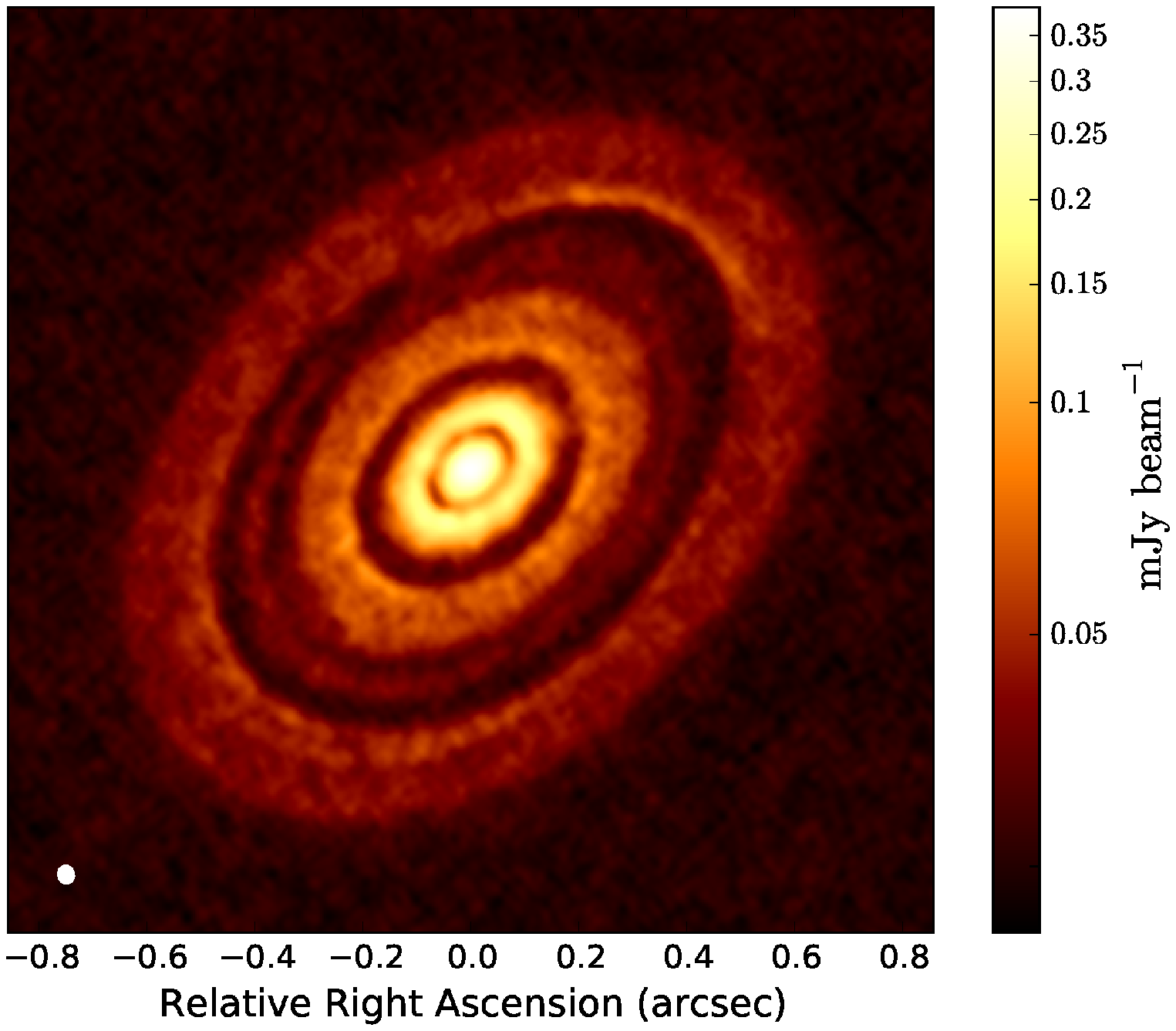}
\caption{Comparison between the ALMA image of HL Tau (left) with simulated observations of our disc model (right) at band 6 (continuum emission at 233 GHz). Note that the color bars are different. The white colour in the filled ellipse in the lower left corner indicates the size of the half-power contour of the synthesized beam: (left) 0.035 arcsec $\times$ 0.022 arcsec, P.A. 11\textdegree; (right) 0.032 arcsec $\times$ 0.027 arcsec, P.A. 12\textdegree.} 
\label{fig:cfr}
\end{center}
\end{figure*}

Fig. \ref{fig:cfr} compares the ALMA simulated observation of our disc model at band 6 (right) with the publicly released image of HL Tau (left). The pattern of bright and dark rings is readily detectable with ALMA using a selection of observation parameters that ensure a resolution close to the one reached in the real ALMA image. As expected, the emission probes the mid-plane disc surface density in large grains (0.1-10 mm), due to their high opacity at these wavelength \citep{dullemondetal07,williamscieza11}. However, the intensity in the ALMA simulated observation of this model is lower than in the HL Tau image. This difference is likely due to the low dust mass, suggesting that the higher disc mass from a steeper surface density profile would produce better signal-to-noise ratio. 

\section{Summary}
\label{sec:discussion}
Inspired by the recent ALMA image of HL Tau, we have performed 3D dust/gas SPH simulations of a disc hosting multiple planets aimed at reproducing the peculiar morphology of alternating dark and bright rings shown in the observation. We used 3D radiative transfer modelling to produce synthetic ALMA observations of our disc model in order to test whether dust features induced by embedded protoplanets can explain those detected. We find:
\begin{enumerate}
\item The absence of spiral structure observed in HL Tau is due to the different response of the dust compared to the gas to embedded planets. Axisymmetric rings such as those observed are a natural product of embedded planets in the disc when observing the dust. Spiral structure may still be present in the gas. Hence dust/gas modelling is crucial for interpreting these systems.
\item In agreement with previous authors, we find it is easier to carve gaps in dust than in gas. Hence gaps observed in the continuum at mm-wavelengths do not necessarily indicate gaps in gas.
\item We can reproduce all the major features of HL Tau observations with three embedded planets at 13.2, 32.3 and 68.8 au with planet masses of 0.2, 0.27 and 0.55 $M_{\rm J}$, respectively, though there is uncertainty in the exact masses from the assumed disc model. This supports the conclusion that embedded protoplanets are responsible. Our masses are similar to the 0.2 $M_{\rm J}$ suggested for all three planets in \citet{dzw14} and consistent with a 10$M_{\rm J}$ upper limit for the third planet given by \citet{testietal15}. 
\item The dust density structures for grains with $S_{\rm t} \gtrsim 10$ (cm-size and larger in our simulations; bottom centre and right panels of Fig. \ref{fig:dust}) show axisymmetric features that can be identified as waves excited by the embedded protoplanets \citep{goldreichtremaine80}. These waves can freely propagate across the dusty disc without being damped by dissipative phenomena \citep{rafikov02}. The lower drag makes these waves more visible since the drag effect on the radial dust motion is negligible. We therefore suggest that the axisymmetric perturbations in the ALMA images in the outer part of the second and third planet orbits are generated by the lower coupling of mm-particles. Further simulations with a steeper surface density profile should be able to reproduce this.
\end{enumerate}

There are many caveats to our results, and these are preliminary investigations. Nevertheless they illustrate that the ALMA observations of HL Tau can be understood from the interaction between dust, gas and planets in the disc.

\section*{Acknowledgements}
GD thanks Monash for hosting and CPU time on NeCTaR from March-July 2015. We thank C.~Dullemond for making RADMC-3D available. We acknowledge an Australian Research Council Future Fellowship and Discovery Project. We used gSTAR, funded by Swinburne University and the Australian Government Education Investment Fund. GD and G. Lodato acknowledge funding via PRIN MIUR 2010-2011 `Chemical and dynamical evolution of the Milky Way and Local Group Galaxies'. G. Laibe is funded by European Research Council FP7 advanced grant ECOGAL. We thank Chris Nixon, Matthew Bate, Leonardo Testi and an anonymous referee for useful discussions.

\label{lastpage}

\bibliography{dan}

\begin{thebibliography}{}
\makeatletter
\relax
\def\mn@urlcharsother{\let\do\@makeother \do\$\do\&\do\#\do\^\do\_\do\%\do\~}
\def\mn@doi{\begingroup\mn@urlcharsother \@ifnextchar [ {\mn@doi@}
  {\mn@doi@[]}}
\def\mn@doi@[#1]#2{\def\@tempa{#1}\ifx\@tempa\@empty \href
  {http://dx.doi.org/#2} {doi:#2}\else \href {http://dx.doi.org/#2} {#1}\fi
  \endgroup}
\def\mn@eprint#1#2{\mn@eprint@#1:#2::\@nil}
\def\mn@eprint@arXiv#1{\href {http://arxiv.org/abs/#1} {{\tt arXiv:#1}}}
\def\mn@eprint@dblp#1{\href {http://dblp.uni-trier.de/rec/bibtex/#1.xml}
  {dblp:#1}}
\def\mn@eprint@#1:#2:#3:#4\@nil{\def\@tempa {#1}\def\@tempb {#2}\def\@tempc
  {#3}\ifx \@tempc \@empty \let \@tempc \@tempb \let \@tempb \@tempa \fi \ifx
  \@tempb \@empty \def\@tempb {arXiv}\fi \@ifundefined
  {mn@eprint@\@tempb}{\@tempb:\@tempc}{\expandafter \expandafter \csname
  mn@eprint@\@tempb\endcsname \expandafter{\@tempc}}}

\bibitem[\protect\citeauthoryear{{ALMA Partnership}}{{ALMA
  Partnership}}{2015}]{almaetal15}
{ALMA Partnership} 2015, arXiv:1503.02649, \href
  {http://adsabs.harvard.edu/abs/2015arXiv150302649P} {}

\bibitem[\protect\citeauthoryear{{Andrews} et~al.}{{Andrews}
  et~al.}{2012}]{andrewsetal12}
{Andrews} S.~M.,  et~al., 2012, \mn@doi [\apj] {10.1088/0004-637X/744/2/162},
  \href {http://adsabs.harvard.edu/abs/2012ApJ...744..162A} {744, 162}

\bibitem[\protect\citeauthoryear{{Ayliffe}, {Laibe}, {Price}  \&
  {Bate}}{{Ayliffe} et~al.}{2012}]{ayliffeetal12}
{Ayliffe} B.~A.,  {Laibe} G.,  {Price} D.~J.,   {Bate} M.~R.,  2012, \mn@doi
  [\mnras] {10.1111/j.1365-2966.2012.20967.x}, \href
  {http://adsabs.harvard.edu/abs/2012MNRAS.423.1450A} {423, 1450}

\bibitem[\protect\citeauthoryear{{Bate}, {Bonnell}  \& {Price}}{{Bate}
  et~al.}{1995}]{bbp95}
{Bate} M.~R.,  {Bonnell} I.~A.,   {Price} N.~M.,  1995, \mnras, \href
  {http://adsabs.harvard.edu/abs/1995MNRAS.277..362B} {277, 362}

\bibitem[\protect\citeauthoryear{{Crida}, {Morbidelli}  \& {Masset}}{{Crida}
  et~al.}{2006}]{cmm06}
{Crida} A.,  {Morbidelli} A.,   {Masset} F.,  2006, \mn@doi [\icarus]
  {10.1016/j.icarus.2005.10.007}, \href
  {http://adsabs.harvard.edu/abs/2006Icar..181..587C} {181, 587}

\bibitem[\protect\citeauthoryear{{Dipierro}, {Pinilla}, {Lodato}  \&
  {Testi}}{{Dipierro} et~al.}{2015}]{dipierroetal15}
{Dipierro} G.,  {Pinilla} P.,  {Lodato} G.,   {Testi} L.,  2015, \mn@doi
  [\mnras] {10.1093/mnras/stv970}, \href
  {http://adsabs.harvard.edu/abs/2015MNRAS.451.5493D} {451, 5493}

\bibitem[\protect\citeauthoryear{{Dong}, {Zhu}  \& {Whitney}}{{Dong}
  et~al.}{2014}]{dzw14}
{Dong} R.,  {Zhu} Z.,   {Whitney} B.,  2014, preprint, \href
  {http://adsabs.harvard.edu/abs/2014arXiv1411.6063D} {} (\mn@eprint {arXiv}
  {1411.6063})

\bibitem[\protect\citeauthoryear{{Draine}}{{Draine}}{2006}]{draine06}
{Draine} B.~T.,  2006, \mn@doi [\apj] {10.1086/498130}, \href
  {http://adsabs.harvard.edu/abs/2006ApJ...636.1114D} {636, 1114}

\bibitem[\protect\citeauthoryear{{Dullemond}}{{Dullemond}}{2012}]{dullemond12}
{Dullemond} C.~P.,  2012, {RADMC-3D: A multi-purpose radiative transfer tool},
  Astrophysics Source Code Library (\mn@eprint {ascl} {1202.015})

\bibitem[\protect\citeauthoryear{{Dullemond} et~al.}{{Dullemond}
  et~al.}{2007}]{dullemondetal07}
{Dullemond} C.~P.,  et~al., 2007, \mn@doi [\aap] {10.1051/0004-6361:20077581},
  \href {http://adsabs.harvard.edu/abs/2007A%26A...473..457D} {473, 457}

\bibitem[\protect\citeauthoryear{{Fouchet} et~al.}{{Fouchet}
  et~al.}{2007}]{fouchetetal07}
{Fouchet} L.,  et~al., 2007, \mn@doi [\aap] {10.1051/0004-6361:20077586}, \href
  {http://adsabs.harvard.edu/abs/2007A%26A...474.1037F} {474, 1037}

\bibitem[\protect\citeauthoryear{{Fouchet}, {Gonzalez}  \&
  {Maddison}}{{Fouchet} et~al.}{2010}]{fgm10}
{Fouchet} L.,  {Gonzalez} J.-F.,   {Maddison} S.~T.,  2010, \mn@doi [\aap]
  {10.1051/0004-6361/200913778}, \href
  {http://adsabs.harvard.edu/abs/2010A%26A...518A..16F} {518, A16}

\bibitem[\protect\citeauthoryear{{Goldreich} \& {Tremaine}}{{Goldreich} \&
  {Tremaine}}{1980}]{goldreichtremaine80}
{Goldreich} P.,  {Tremaine} S.,  1980, \mn@doi [\apj] {10.1086/158356}, \href
  {http://adsabs.harvard.edu/abs/1980ApJ...241..425G} {241, 425}

\bibitem[\protect\citeauthoryear{{Greaves} et~al.}{{Greaves}
  et~al.}{2008}]{greavesetal08}
{Greaves} J.~S.,  et~al., 2008, \mn@doi [\mnras]
  {10.1111/j.1745-3933.2008.00559.x}, \href
  {http://adsabs.harvard.edu/abs/2008MNRAS.391L..74G} {391, L74}

\bibitem[\protect\citeauthoryear{{Hayashi}, {Ohashi}  \& {Miyama}}{{Hayashi}
  et~al.}{1993}]{hom93}
{Hayashi} M.,  {Ohashi} N.,   {Miyama} S.~M.,  1993, \mn@doi [\apjl]
  {10.1086/187119}, \href {http://adsabs.harvard.edu/abs/1993ApJ...418L..71H}
  {418, L71}

\bibitem[\protect\citeauthoryear{{Kwok}}{{Kwok}}{1975}]{kwok75}
{Kwok} S.,  1975, \mn@doi [\apj] {10.1086/153637}, \href
  {http://adsabs.harvard.edu/abs/1975ApJ...198..583K} {198, 583}

\bibitem[\protect\citeauthoryear{{Kwon}, {Looney}  \& {Mundy}}{{Kwon}
  et~al.}{2011}]{klm11}
{Kwon} W.,  {Looney} L.~W.,   {Mundy} L.~G.,  2011, \mn@doi [\apj]
  {10.1088/0004-637X/741/1/3}, \href
  {http://adsabs.harvard.edu/abs/2011ApJ...741....3K} {741, 3}

\bibitem[\protect\citeauthoryear{{Kwon} et~al.}{{Kwon}
  et~al.}{2015}]{kwonetal15}
{Kwon} W.,  et~al., 2015, arXiv:1506.03679, \href
  {http://adsabs.harvard.edu/abs/2015arXiv150603679K} {}

\bibitem[\protect\citeauthoryear{{Laibe} \& {Price}}{{Laibe} \&
  {Price}}{2011}]{laibeprice11}
{Laibe} G.,  {Price} D.~J.,  2011, \mn@doi [\mnras]
  {10.1111/j.1365-2966.2011.19291.x}, \href
  {http://adsabs.harvard.edu/abs/2011MNRAS.418.1491L} {418, 1491}

\bibitem[\protect\citeauthoryear{{Laibe} \& {Price}}{{Laibe} \&
  {Price}}{2012a}]{laibeprice12}
{Laibe} G.,  {Price} D.~J.,  2012a, \mn@doi [\mnras]
  {10.1111/j.1365-2966.2011.20202.x}, \href
  {http://adsabs.harvard.edu/abs/2012MNRAS.420.2345L} {420, 2345}

\bibitem[\protect\citeauthoryear{{Laibe} \& {Price}}{{Laibe} \&
  {Price}}{2012b}]{laibeprice12a}
{Laibe} G.,  {Price} D.~J.,  2012b, \mn@doi [\mnras]
  {10.1111/j.1365-2966.2011.20201.x}, \href
  {http://adsabs.harvard.edu/abs/2012MNRAS.420.2365L} {420, 2365}

\bibitem[\protect\citeauthoryear{{Laibe} \& {Price}}{{Laibe} \&
  {Price}}{2014}]{laibeprice14}
{Laibe} G.,  {Price} D.~J.,  2014, \mn@doi [\mnras] {10.1093/mnras/stu355},
  \href {http://adsabs.harvard.edu/abs/2014MNRAS.440.2136L} {440, 2136}

\bibitem[\protect\citeauthoryear{{Lin} \& {Papaloizou}}{{Lin} \&
  {Papaloizou}}{1986}]{linpapaloizou86}
{Lin} D.~N.~C.,  {Papaloizou} J.,  1986, \mn@doi [\apj] {10.1086/164426}, \href
  {http://adsabs.harvard.edu/abs/1986ApJ...307..395L} {307, 395}

\bibitem[\protect\citeauthoryear{{Lodato} \& {Price}}{{Lodato} \&
  {Price}}{2010}]{lodatoprice10}
{Lodato} G.,  {Price} D.~J.,  2010, \mn@doi [\mnras]
  {10.1111/j.1365-2966.2010.16526.x}, \href
  {http://adsabs.harvard.edu/abs/2010MNRAS.405.1212L} {405, 1212}

\bibitem[\protect\citeauthoryear{{Lodato} \& {Rice}}{{Lodato} \&
  {Rice}}{2004}]{lodatorice04}
{Lodato} G.,  {Rice} W.~K.~M.,  2004, \mn@doi [\mnras]
  {10.1111/j.1365-2966.2004.07811.x}, \href
  {http://adsabs.harvard.edu/abs/2004MNRAS.351..630L} {351, 630}

\bibitem[\protect\citeauthoryear{{Lyra} \& {Kuchner}}{{Lyra} \&
  {Kuchner}}{2013}]{lyrakuchner13}
{Lyra} W.,  {Kuchner} M.,  2013, \mn@doi [\nat] {10.1038/nature12281}, \href
  {http://adsabs.harvard.edu/abs/2013Natur.499..184L} {499, 184}

\bibitem[\protect\citeauthoryear{{Maddison}, {Fouchet}  \&
  {Gonzalez}}{{Maddison} et~al.}{2007}]{mfg07}
{Maddison} S.~T.,  {Fouchet} L.,   {Gonzalez} J.-F.,  2007, \mn@doi [\apss]
  {10.1007/s10509-007-9572-y}, \href
  {http://adsabs.harvard.edu/abs/2007Ap%26SS.311....3M} {311, 3}

\bibitem[\protect\citeauthoryear{{Mundt}, {Buehrke}, {Solf}, {Ray}  \&
  {Raga}}{{Mundt} et~al.}{1990}]{mundtetal90}
{Mundt} R.,  {Buehrke} T.,  {Solf} J.,  {Ray} T.~P.,   {Raga} A.~C.,  1990,
  \aap, \href {http://adsabs.harvard.edu/abs/1990A%26A...232...37M} {232, 37}

\bibitem[\protect\citeauthoryear{{Nixon}, {King}  \& {Price}}{{Nixon}
  et~al.}{2013}]{nkp13}
{Nixon} C.,  {King} A.,   {Price} D.,  2013, \mn@doi [\mnras]
  {10.1093/mnras/stt1136}, \href
  {http://adsabs.harvard.edu/abs/2013MNRAS.434.1946N} {434, 1946}

\bibitem[\protect\citeauthoryear{{Paardekooper} \& {Mellema}}{{Paardekooper} \&
  {Mellema}}{2004}]{paardekoopermellema04}
{Paardekooper} S.-J.,  {Mellema} G.,  2004, \mn@doi [\aap]
  {10.1051/0004-6361:200400053}, \href
  {http://adsabs.harvard.edu/abs/2004A%26A...425L...9P} {425, L9}

\bibitem[\protect\citeauthoryear{{Paardekooper} \& {Mellema}}{{Paardekooper} \&
  {Mellema}}{2006}]{paardekoopermellema06}
{Paardekooper} S.-J.,  {Mellema} G.,  2006, \mn@doi [\aap]
  {10.1051/0004-6361:20054449}, \href
  {http://adsabs.harvard.edu/abs/2006A%26A...453.1129P} {453, 1129}

\bibitem[\protect\citeauthoryear{{Pinilla} et~al.}{{Pinilla}
  et~al.}{2012}]{pinillaetal12}
{Pinilla} P.,  et~al., 2012, \mn@doi [\aap] {10.1051/0004-6361/201118204},
  \href {http://adsabs.harvard.edu/abs/2012A%26A...538A.114P} {538, A114}

\bibitem[\protect\citeauthoryear{{Price}}{{Price}}{2012}]{price12}
{Price} D.~J.,  2012, \mn@doi [J. Comp. Phys.] {10.1016/j.jcp.2010.12.011},
  \href {http://adsabs.harvard.edu/abs/2012JCoPh.231..759P} {231, 759}

\bibitem[\protect\citeauthoryear{{Price} \& {Federrath}}{{Price} \&
  {Federrath}}{2010}]{pricefederrath10}
{Price} D.~J.,  {Federrath} C.,  2010, \mn@doi [\mnras]
  {10.1111/j.1365-2966.2010.16810.x}, \href
  {http://adsabs.harvard.edu/abs/2010MNRAS.406.1659P} {406, 1659}

\bibitem[\protect\citeauthoryear{{Price} \& {Laibe}}{{Price} \&
  {Laibe}}{2015}]{pricelaibe15}
{Price} D.~J.,  {Laibe} G.,  2015, {\mnras}, \href
  {http://adsabs.harvard.edu/abs/2015MNRAS.451.5332P} {451, 5332}

\bibitem[\protect\citeauthoryear{{Rafikov}}{{Rafikov}}{2002}]{rafikov02}
{Rafikov} R.~R.,  2002, \mn@doi [\apj] {10.1086/340228}, \href
  {http://adsabs.harvard.edu/abs/2002ApJ...572..566R} {572, 566}

\bibitem[\protect\citeauthoryear{{Robitaille} et~al.}{{Robitaille}
  et~al.}{2007}]{robitailleetal07}
{Robitaille} T.~P.,  et~al., 2007, \mn@doi [\apjs] {10.1086/512039}, \href
  {http://adsabs.harvard.edu/abs/2007ApJS..169..328R} {169, 328}

\bibitem[\protect\citeauthoryear{{Shakura} \& {Sunyaev}}{{Shakura} \&
  {Sunyaev}}{1973}]{shakurasunyaev73}
{Shakura} N.~I.,  {Sunyaev} R.~A.,  1973, \aap, \href
  {http://adsabs.harvard.edu/abs/1973A%26A....24..337S} {24, 337}

\bibitem[\protect\citeauthoryear{{Takeuchi} \& {Lin}}{{Takeuchi} \&
  {Lin}}{2002}]{takeuchilin02}
{Takeuchi} T.,  {Lin} D.~N.~C.,  2002, \mn@doi [\apj] {10.1086/344437}, \href
  {http://adsabs.harvard.edu/abs/2002ApJ...581.1344T} {581, 1344}

\bibitem[\protect\citeauthoryear{{Tamayo}, {Triaud}, {Menou}  \&
  {Rein}}{{Tamayo} et~al.}{2015}]{tamayoetal15}
{Tamayo} D.,  {Triaud} A.~H.~M.~J.,  {Menou} K.,   {Rein} H.,  2015, \mn@doi
  [\apj] {10.1088/0004-637X/805/2/100}, \href
  {http://adsabs.harvard.edu/abs/2015ApJ...805..100T} {805, 100}

\bibitem[\protect\citeauthoryear{{Testi} et~al.}{{Testi}
  et~al.}{2015}]{testietal15}
{Testi} L.,  et~al., 2015, \apj, submitted

\bibitem[\protect\citeauthoryear{{Weidenschilling}}{{Weidenschilling}}{1977}]{weidenschilling77}
{Weidenschilling} S.~J.,  1977, \mnras, \href
  {http://adsabs.harvard.edu/abs/1977MNRAS.180...57W} {180, 57}

\bibitem[\protect\citeauthoryear{{Williams} \& {Cieza}}{{Williams} \&
  {Cieza}}{2011}]{williamscieza11}
{Williams} J.~P.,  {Cieza} L.~A.,  2011, \mn@doi [\araa]
  {10.1146/annurev-astro-081710-102548}, \href
  {http://adsabs.harvard.edu/abs/2011ARA%26A..49...67W} {49, 67}

\bibitem[\protect\citeauthoryear{{Wolf}, {Gueth}, {Henning}  \& {Kley}}{{Wolf}
  et~al.}{2002}]{wolfetal02}
{Wolf} S.,  {Gueth} F.,  {Henning} T.,   {Kley} W.,  2002, \mn@doi [\apjl]
  {10.1086/339544}, \href {http://adsabs.harvard.edu/abs/2002ApJ...566L..97W}
  {566, L97}

\bibitem[\protect\citeauthoryear{{Zhang}, {Blake}  \& {Bergin}}{{Zhang}
  et~al.}{2015}]{zbb15}
{Zhang} K.,  {Blake} G.~A.,   {Bergin} E.~A.,  2015, \mn@doi [\apjl]
  {10.1088/2041-8205/806/1/L7}, \href
  {http://adsabs.harvard.edu/abs/2015ApJ...806L...7Z} {806, L7}

\bibitem[\protect\citeauthoryear{{de Gregorio-Monsalvo} et~al.}{{de
  Gregorio-Monsalvo} et~al.}{2013}]{de-gregorio-monsalvoetal13}
{de Gregorio-Monsalvo} I.,  et~al., 2013, \mn@doi [\aap]
  {10.1051/0004-6361/201321603}, \href
  {http://adsabs.harvard.edu/abs/2013A%26A...557A.133D} {557, A133}

\bibitem[\protect\citeauthoryear{{de Val-Borro} et~al.}{{de Val-Borro}
  et~al.}{2006}]{de-val-borroetal06}
{de Val-Borro} M.,  et~al., 2006, \mn@doi [\mnras]
  {10.1111/j.1365-2966.2006.10488.x}, \href
  {http://adsabs.harvard.edu/abs/2006MNRAS.370..529D} {370, 529}

\makeatother
\end{thebibliography}

\end{document}